\documentclass[lettersize,journal]{IEEEtran}
\usepackage{amsmath,amsfonts}
\usepackage{algorithmic}
\usepackage{algorithm}
\usepackage{array}
\usepackage[caption=false,font=normalsize,labelfont=sf,textfont=sf]{subfig}
\usepackage{textcomp}
\usepackage{stfloats}
\usepackage{url}
\usepackage{verbatim}
\usepackage{graphicx}
\usepackage{cite}
\usepackage[mathscr]{euscript}
\usepackage{booktabs}
\usepackage{bbm}
\usepackage{amssymb}
\usepackage{glossaries}
\usepackage{bm}
\usepackage{setspace} 
\newacronym{AN}{AN}{artificial noise}
\newacronym{ISAC}{ISAC}{Integrated sensing and communications}
\newacronym{MI}{MI}{mutual information}
\newacronym{SINR}{SINR}{signal-to-interference-plus-noise ratio}
\newacronym{CU}{CU}{communication user}
\newacronym{BS}{BS}{base station}
\newacronym{DFRC}{DFRC}{dual-functional radar-communication}
\newacronym{CRB}{CRB}{Cram\'{e}r-Rao bound}
\newacronym{SNR}{SNR}{signal-to-noise ratio}
\newacronym{SCA}{SCA}{successive convex approximation}
\newacronym{SDR}{SDR}{semidefinite relaxation}
\newacronym{MIMO}{MIMO}{multiple-input multiple-output}
\newacronym{Eve}{Eve}{eavesdropper}
\newacronym{LoS}{LoS}{line-of-sight}

\begin{document}

\title{Securing the Sensing Functionality in ISAC Networks: An Artificial Noise Design}
\author{Jiaqi Zou,~\IEEEmembership{Graduate Student Member,~IEEE}, Christos Masouros,~\IEEEmembership{Senior Member,~IEEE},  Fan Liu,~\IEEEmembership{Member,~IEEE} Songlin Sun,~\IEEEmembership{Senior Member,~IEEE}
\thanks{Jiaqi Zou is with the School of Information and Communication Engineering, Beijing University of Posts and Telecommunications (BUPT), Beijing, 100876, China, and also with the Department of Electrical and Electronic Engineering, University College London, London, WC1E 7JE, UK (e-mail: jqzou@bupt.edu.cn). }
\thanks{Christos Masouros is with the Department of Electronic and Electrical Engineering, University College London, London, WC1E 7JE, UK (e-mail: chris.masouros@ieee.org).}
\thanks{Fan Liu is with the Department of Electrical and Electronic Engineering, Southern University of Science and Technology, Shenzhen 518055, China (e-mail: liuf6@sustech.edu.cn).}
\thanks{Songlin Sun is with Beijing University of Posts and Telecommunications (BUPT), Beijing, 100876, China (e-mail: slsun@bupt.edu.cn).}
}

\markboth{}%
{Shell \MakeLowercase{\textit{et al.}}: A Sample Article Using IEEEtran.cls for IEEE Journals}


\maketitle

\begin{abstract}
Integrated sensing and communications (ISAC) systems employ dual-functional signals to simultaneously accomplish radar sensing and wireless communication tasks. However, ISAC systems open up new \textit{sensing security} vulnerabilities to malicious illegitimate eavesdroppers (Eves) that can also exploit the transmitted waveform to extract sensing information from the environment. In this paper, we investigate the beamforming design to enhance the sensing security of an ISAC system, where the communication user (CU) serves as a sensing Eve. Our objective is to maximize the mutual information (MI) for the legitimate radar sensing receiver while considering the constraint of the MI for the Eve and the quality of service to the CUs. Then, we consider the artificial noise (AN)-aided beamforming to further enhance the sensing security. Simulation results demonstrate that our proposed methods achieve MI improvement of the legitimate receiver while limiting the sensing MI of the Eve, compared with the baseline scheme, and that the utilization of AN further contributes to sensing security.
\end{abstract}

\begin{IEEEkeywords}
Integrated sensing and communications, sensing security, mutual information, artificial noise.
\end{IEEEkeywords}
\vspace{-2mm}
\section{Introduction}

\gls{ISAC} is identified as a key 6G technology that will support various futuristic applications through the co-design of sensing and communication functionalities.  
In particular, supported by the implementation of the dual-functional waveform and the \glspl{BS}, ISAC provides a step change from the spectral coexistence of radar and communication systems to the shared utilization of costly hardware platforms. 
Inspired by these favorable characteristics, various designs have been proposed for the dual-functional waveforms to promote sensing and communication performance. For example, recent works in \cite{hua2023optimal} minimized the beampattern matching error for radar sensing, and \cite{liu2021cram} considered the \gls{CRB} minimization subject to the minimum \gls{SINR} constraints for each \gls{CU}.  However, the aforementioned works have overlooked the consideration of security issues, which avail unique vulnerabilities in ISAC systems.

Due to the inherent broadcast nature of wireless signals, it is inevitable that wireless communication/sensing systems are susceptible to potential security threats. Compared with the communication-only systems, ISAC systems encounter more intricate security issues which can be generally categorized into the information security for communication and the sensing security for radar. The former arises from the fact that the probing ISAC waveform is modulated with information, which could potentially be leaked to the sensed targets that can act as \glspl{Eve}. To deal with this, physical layer security schemes have been proposed, such as \cite{xu2022robust} that maximized the secrecy rate by jointly optimizing the beamforming vector, the duration of snapshots, and the covariance matrix of the \gls{AN}.
Besides, the work in \cite{su2020secure} optimizes the beamforming and \gls{AN} design to minimize the \gls{SNR} at the \gls{Eve}. In contrast, the sensing security for radar has not been well investigated.
In particular, the transmitted waveform could be exploited by the communication user serving as malicious passive sensing \gls{Eve}s to extract sensing information about targets or their environment, which poses significant challenges and the need to secure the sensing functionality of ISAC networks.

Against this background, our work proposes to address the sensing security issue for  ISAC systems, which to the best of our knowledge, has not been investigated in the literature. In particular, we consider a bi-static ISAC scenario, where the users that are granted access only to communication services, can exploit the ISAC signals for illegitimate sensing and therefore take the role of sensing Eves. In this scenario, the legitimate receiver and the \gls{Eve} conduct sensing with the dual-functional ISAC waveforms. We formulate the problem to maximize the radar \gls{MI} of the legitimate receiver, subject to MI of the \gls{Eve}, the minimum \gls{SINR} constraints for each \gls{CU} and the maximum transmit power budget. Since the formulated problem is nonconvex, we propose a \gls{SCA} method combined with \gls{SDR} to deal with the non-convexity. Furthermore, we propose an \gls{AN}-aided method where we firstly give the derivation of sensing MI with AN and then jointly optimize the beamforming and the covariance of AN. Simulation results demonstrate significant improvement in the sensing MI of our proposed methods compared with the baseline and also reveal that through adding AN to the transmit signals of the BS, the secure sensing performance can be effectively improved.

\vspace{-2mm}
\section{System Model}
\begin{figure}
	\centering
	\includegraphics[width=0.8\linewidth]{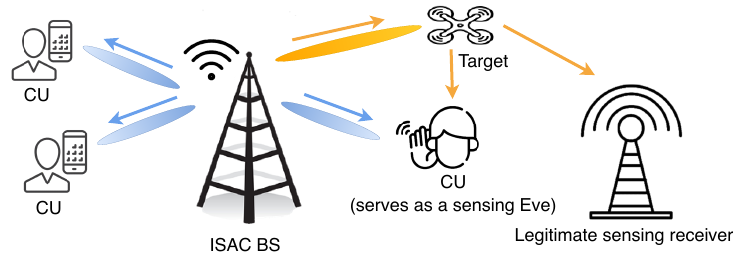}
	\caption{System model.}
	\label{fig:systemmodel}
\end{figure}

We consider a bistatic \gls{MIMO} ISAC system, which consists of a central \gls{BS} transmitting dual-functional signals to a legitimate radar receiver and $K$ single-antenna \glspl{CU}. Simultaneously, a CU serves as an unauthorized sensing \gls{Eve} who perfectly knows/intercepts the transmitted signals and also wishes to sense the targets/environment. We assume that the transmitter is equipped with a uniform linear array (ULA) of $N_t$ antennas and that the legitimate receiver and the \gls{Eve} are equipped with $N_r$ and $N_e$ antennas, respectively.



\vspace{-2mm}
\subsection{Communication Signal Model and Metrics}

Let ${\bf h}_k \in \mathbb{C}^{N_t\times 1}$ represent the communication channel vector for the $k$-th use. We assume that the channel follows a slow-fading block Rician fading channel, given as 
\begin{align}
\mathbf{h}_k = 	\sqrt{ \frac{K_k}{K_k + 1}} \mathbf{h}_{LoS, k} +  \sqrt{ \frac{1}{K_k + 1}} \mathbf{h}_{NLoS, k},
\end{align} 
where $K_k$ denotes the Rician factor of the channel between the $k$-th CU and the BS. $\mathbf{h}_{LoS, k}$ denotes the deterministic LoS channel component with $\mathbf{h}_{LoS, k}= \left[ 1, \cdots, e^{-j \pi (N_t -1) cos\theta_k} \right] $, where $\theta_k \in [0, \pi]$ is the angle-of-arrival (AoA) of the \gls{LoS} link from the $k$-th CU to the BS and we assume half-wavelength antenna spacing. $\mathbf{h}_{NLoS, k}$ represents the random scattered components whose elements follows $\mathcal{CN}(0,1)$.

Let us denote the beamforming matrix as $\mathbf{W} = [\mathbf{w}_1, \mathbf{w}_2, \cdots, \mathbf{w}_k]  \in \mathcal{C}^{N_t \times K}$, where ${\bf w}_k \in \mathbb{C}^{N_t\times 1}$  stands for the beamforming vector of the $k$-th user. Then, the transmitted signal at the $l$-th time slot can be expressed as 
\begin{align}
\mathbf{x}[l]= \mathbf{Ws}[l],
\end{align}
\noindent where $\mathbf{s}[l] \in \mathcal{C}^{K \times 1}$ includes $K$ parallel communication symbol streams to be communicated to $K$ users. Without loss of generality, we assume the communication symbols have unit power, i.e., $\mathbb{E}[\mathbf{s}[l] \mathbf{s}[l]^H] = \mathbf{I}$, and the communication channel matrix is perfectly estimated and known at the BS side. Then, we have the SINR at the $k$-th CU as
%
%
\begin{align}
	\text{SINR}_k = \frac{\left\vert {\bf h}_k^{H} {\bf w}_k \right\vert^2}{\sigma_c^2 + \sum_{j=1, j \neq k}^K \left\vert {\bf h}_k^H {\bf w}_j \right\vert^2}, \label{eq:sinr}
\end{align}
where $\sigma_c^2$ is the variance of additive white Gaussian noise (AWGN). 


\subsection{Radar Signal Model}
We consider a bistatic radar sensing scenario where the transmitted signals are shared between the transmitter and receiver through a control center. Denoting $\mathbf{X} = [\mathbf{x}[1], \mathbf{x}[2], \cdots, \mathbf{x}[L]] \in \mathcal{C}^{N_t \times L} $ as the transmitted waveform during $L$ time slots, the received signal matrix at the legitimate receiver and the \gls{Eve} can be expressed, respectively, as 
\begin{align}
	\mathbf{Y}_{r} = \mathbf{H}_{r}(\bm{\theta}_r) \mathbf{X} + \mathbf{Z}_{r}, 
	\mathbf{Y}_{e} = \mathbf{H}_{e}(\bm{\theta}_e) \mathbf{X} + \mathbf{Z}_{e}, 
\end{align}
where $\mathbf{Z}_{r} $ and $ \mathbf{Z}_{e}$ are the noise matrices at the legitimate receiver and the \gls{Eve}, respectively and each columns of  $\mathbf{Z}_{r} $ and $\mathbf{Z}_{e} $ follow $\mathcal{CN}(\mathbf{0}, \sigma_{r}^2 \mathbf{I})$ and $\mathcal{CN}(\mathbf{0}, \sigma_{e}^2 \mathbf{I})$, respectively. $\mathbf{H}_{r}$ and $\mathbf{H}_{e}$ are the target response matrices from the transmitter to the legitimate receiver and the \gls{Eve}, respectively, which are determined by the target parameters of interests $\bm{\theta}_r$ and $\bm{\theta}_e$. Note that $\bm{\theta}_r \neq \bm{\theta}_e$ since the legitimate receiver and the \gls{Eve} may have different sensing interests to estimate different angles and ranges, etc. Aside of $\mathbf{H}_r$, we note that $\mathbf{H}_e$ can also be known at the BS by exploiting the knowledge of the Eve's location which is typical information exchanged with the BS, since the Eve is also a legitimate CU.

To measure the sensing performance at the receivers, we adopt the mutual information between the received radar signal and target response matrix as the performance metric for radar sensing, which has been extensively discussed in the literature~\cite{yang2007mimo}. It's also implied in~\cite{bell1993information} that the MI characterizes the capability of radar to estimate the parameters describing the target. Following~\cite{liu2022deterministic}, the sensing \gls{MI} of the legitimate receiver can be given as
\begin{align}
		& I_r(\mathbf{Y}_{r}; \bm{\theta}_r | \mathbf{X}) =  I_r(\mathbf{Y}_{r}; \mathbf{H}_{r} | \mathbf{X}).
\end{align}
Vectorizing $\mathbf{Y}$, we have
\begin{align}
\mathrm{vec}(\mathbf{Y}_r) =\tilde{\mathbf{X}} \mathbf{h}_r + \mathbf{z}_r,
\end{align}
where $\tilde{\mathbf{X}} = \mathbf{X}^T \otimes \mathbf{I}_{N_r}$, $\mathbf{h}_r = \mathrm{vec}(\mathbf{H}_r)$, and $\mathbf{z}_r = \mathrm{vec}(\mathbf{Z}_r)$\cite[Eq. 1.11.20]{zhang2017matrix}. Following the assumptions are also used in \cite{yang2007mimo}, we assume the target response vector $\mathbf{h}_r$ is zero-mean circular-symmetric
Gaussian distributed with covariance $\mathbf{R}_h$.
As the transmitted waveform is perfectly known, the \gls{MI} between $\mathbf{Y}_r$ and $\mathbf{H}_r$ at the legitimate receiver can be expressed as
\begin{subequations}
\begin{align}
	I_r(\mathbf{Y}_{r}; \mathbf{H}_{r} | \mathbf{X}) &= h(\mathbf{Y}_{r}|  \mathbf{X}) - h(\mathbf{Y}_{r} | \mathbf{H}_r,  \mathbf{X}) \\
	&= \log \det( \mathbf{I} + \sigma_{r}^{-2} \tilde{\mathbf{X}} \mathbf{R}_{h_r} \tilde{\mathbf{X}}^H) \\
	&=\log \det( \mathbf{I} + \sigma_{r}^{-2}   \mathbf{R}_{h_r}  (\mathbf{X}^\ast \mathbf{X}^T\otimes \mathbf{I}_{N_r})) ,
\end{align}
\end{subequations}
where we used the property of matrix determinant that $\mathrm{det}(\mathbf{I} + \mathbf{AB}) = \mathrm{det}(\mathbf{I} + \mathbf{BA}) $. Let $\mathbf{R}_{h_r} = \mathbf{U}_r \mathbf{\Lambda}_r \mathbf{U}_r^H$ be the eigenvalue decomposition of $\mathbf{R}_{h_r}$, and $\mathbf{K} $ be a real commutation matrix satisfying $\mathbf{K} \mathbf{K}^T = \mathbf{I}$. We then have
\begin{align}
		I_r &= \log \det( \mathbf{I} + \sigma_{r}^{-2} \mathbf{\Lambda}_r \mathbf{U}_r^H \mathbf{K}   (  \mathbf{I}_{N_r} \otimes \mathbf{X}^\ast \mathbf{X}^T ) \mathbf{K}^T \mathbf{U}_r) \notag \\
		&= \log \det (\mathbf{I} +  \sigma_{r}^{-2} L \mathbf{\Lambda}_r \sum_{i=1}^{N_r} \mathbf{P}_i \mathbf{R}_\mathbf{X}^\ast \mathbf{P}_i^H),
	\end{align}
where $\mathbf{P} = \mathbf{U}_r^H \mathbf{K} = [\mathbf{P}_{1}, \mathbf{P}_2, \cdots, \mathbf{P}_{N_r}]$ and $\mathbf{R}_\mathbf{X}$ denotes the covariance matrix of the transmit signal, given as ~\cite{liu2021cram}
\begin{align}
	{\bf R}_\mathbf{X} = \frac{1}{L}{\bf X}{\bf X}^H \approx {\bf W} {\bf W}^H = \sum_{k=1}^K {\bf w}_k{\bf w}_k^H,
\end{align}
where the approximation holds when $L$ is large enough. Similarly, the MI between $\mathbf{Y}_r$ and $\mathbf{H}_r$ at the \gls{Eve} can be expressed as
\begin{align} 
	I_e(\mathbf{Y}_{e}; \mathbf{H}_{e} | \mathbf{X}) \!\!&= \log \det( \mathbf{I} + \sigma_{e}^{-2}   \mathbf{R}_{h_e}  (\mathbf{X}^\ast \mathbf{X}^T\otimes \mathbf{I}_{N_e})) \notag \\
	&= \log \det (\mathbf{I} +  \sigma_{e}^{-2} L \mathbf{\Lambda}_e \sum_{i=1}^{N_e} {\mathbf{Q}}_i \mathbf{R}_\mathbf{X}^\ast \mathbf{Q}_i^H), \label{eq:eveMI}
\end{align}
where $ \mathbf{R}_{h_e}$ denotes the covariance of zero-mean circular-symmetric Gaussian distributed $\mathbf{h}_e = \mathrm{vec}(\mathbf{H}_e)$, whose eigenvalue decomposition is given as $\mathbf{R}_{h_e} = \mathbf{U}_e \mathbf{\Lambda}_e \mathbf{U}_e^H$.  ${\mathbf{Q}} = \mathbf{U}_e^H \mathbf{K} = [{\mathbf{Q}}_{1}, {\mathbf{Q}}_2, \cdots, {\mathbf{Q}}_{N_e}]$.
\section{Beamforming Design for Sensing Security without AN} \label{sec:withoutAN}

In this section, we first investigate the beamforming design without the aid of \gls{AN} to guarantee secure sensing.
Our objective is to maximize the MI of the legitimate receiver while keeping the MI of the \gls{Eve} lower than the preset threshold, and satisfying multiple users’ required SINR and the power budget. The optimization problem can be formulated as follows
\begin{subequations} \label{eq:p1-1}
\begin{align} 
		\max_{\left\lbrace \mathbf{W}_k\right\rbrace_{k=1}^K } &  I_r = \log \det \left( \mathbf{I} +  \sigma_{r}^{-2} L \mathbf{\Lambda}_r \sum_{i=1}^{N_r} \mathbf{P}_i \mathbf{R}_\mathbf{X}^\ast \mathbf{P}_i^H\right)  \\
		\mathrm{s.t.}~& I_e = \log \det \left( \mathbf{I} +  \sigma_{e}^{-2} L \mathbf{\Lambda}_e \sum_{i=1}^{N_e} \mathbf{Q}_i \mathbf{R}_\mathbf{X}^\ast \mathbf{Q}_i^H\right)  \leq \epsilon, \label{eq:p1b} \\
		&\text{tr}(\mathbf{R}_\mathbf{X} )  \leq P_0, \label{eq:p1c}\\
		& \frac{\text{tr}\left(\mathbf{h}_k  \mathbf{W}_k \mathbf{h}_k^H \right)}{\sum\nolimits_{k=1, k\neq i}^{K} \text{tr}\left(\mathbf{h}_k  \mathbf{W}_i \mathbf{h}_k^H \right)+\sigma_c^{2}} \geq \gamma_k, \forall k, \label{eq:p1d}\\
		& \mathbf{R}_\mathbf{X} = \! \sum_{k=1}^K \mathbf{W}_k, \mathbf{W}_k \! \succeq \! 0, \text{rank}(\mathbf{W}_k) = 1, \forall k, \label{eq:p1e}
\end{align} %
\end{subequations} 
where $\mathbf{W}_k = \mathbf{w}_k \mathbf{w}_k^H$.
In general, it is challenging to solve problem \eqref{eq:p1-1} directly, due to the nonconvexity of the constraint \eqref{eq:p1b}, \eqref{eq:p1d}, and the rank-1 constraint in  \eqref{eq:p1e}.
For addressing the nonconvex constraint \eqref{eq:p1b}, we notice that $I_e$ is a concave function in $\mathbf{R}_\mathbf{X}$. Thus,
we give the upper-bound of \eqref{eq:p1b} based on first-order Taylor expansion at a given transmit covariance matrix $\tilde{\mathbf{R}}_\mathbf{X} = \sum_{k=1}^K \tilde{\mathbf{W}}_k$ as 
\begin{align}
\tilde{I}_{e} \triangleq & f(\tilde{\mathbf{R}}_\mathbf{X} ) + \mathrm{tr}\left( \mathrm{Re}\left( 2 \sigma_{e}^{-2} L  \sum_{i=1}^{N_e}  \mathbf{Q}_i^T  (\mathbf{M}^{-1})^\ast  \mathbf{\Lambda}_e  \mathbf{Q}_i^\ast \mathbf{R_X}  \right)  \right) \notag \\
	&-  \mathrm{tr}\left( \mathrm{Re}\left( 2 \sigma_{e}^{-2} L  \sum_{i=1}^{N_e}  \mathbf{Q}_i^T  (\mathbf{M}^{-1})^\ast  \mathbf{\Lambda}_e  \mathbf{Q}_i^\ast  \tilde{\mathbf{R}}_\mathbf{X}   \right)  \right).
\end{align}
where $\mathbf{M} = \mathbf{I} +  \sigma_{e}^{-2} L \mathbf{\Lambda}_e \sum_{i=1}^{N_e} \mathbf{Q}_i \tilde{\mathbf{R}}_X^\ast \mathbf{Q}_i^H$. The details are given in Appendix A. As such, it can be easily observed that the approximated function is convex on $\mathbf{R}_\mathbf{X}$. Thus, $\mathbf{R}_\mathbf{X}$ can be iteratively obtained by updating $\tilde{\mathbf{R}}_\mathbf{X} $. 

Then, we focus on dealing with the rank-1 constraint in \eqref{eq:p1e}, which can be equivalently transformed into an equivalent linear matrix inequality (LMI) as
\begin{subequations}
	\begin{align} 
		& \begin{bmatrix}
			\mathbf{W}_k      & \mathbf{w}_k       \\
			\mathbf{w}_k^H     & 1    
		\end{bmatrix} \succeq 0 ,   \label{eq:rank1a} \\
		& \operatorname{tr}(\mathbf{W}_k) - \mathbf{w}^H_k \mathbf{w}_k \leq 0, \forall k. \label{eq:rank1b}
	\end{align}
\end{subequations}

Additionally, the non-convexity in \eqref{eq:rank1b} can be handled by the first-order Taylor expansions at $\tilde{\mathbf{w}}_k$ as 
\begin{align}
	\operatorname{tr}(\mathbf{W}_k) - {\tilde{\mathbf{w}}_k}^H \tilde{\mathbf{w}}_k - 2\operatorname{Re}\left(\tilde{\mathbf{w}}_k^H \mathbf{w}_k \right) \leq 0,  \label{eq:rank1bconvex}
\end{align}
where $\operatorname{Re}(\cdot)$ denotes the real part of the argument and $\tilde{\mathbf{w}}_k$ can be updated at each iteration. Therefore, a convex approximation of problem \eqref{eq:p1-1} is reformulated as
\begin{subequations}  \label{eq:p1-2}
	\begin{align} 
	& \max_{\left\lbrace  \mathbf{w}_k, \mathbf{W}_k\right\rbrace_{k=1}^K }  I_r	\\
	s.t.~& \tilde{I}_e \leq \epsilon, \label{eq:p1-2b} \\
	& \text{tr}( \mathbf{R}_\mathbf{X})  \leq P_0, \label{eq:p1-power} \\
	& {\bf W}_k \succeq 0, \mathbf{R}_\mathbf{X} = \! \sum_{k=1}^K \mathbf{W}_k,  \label{eq:p1-2c} \\
	& \text{tr}\left(\mathbf{h}_k  \mathbf{W}_k \mathbf{h}_k^H \right)- \gamma_k \! \! \!\!\! \sum_{k\neq i, k=1}^{K} \!\!\!\!\! \text{tr}\left(\mathbf{h}_k  \mathbf{W}_k \mathbf{h}_k^H \right)\! \geq\! \gamma_k \sigma_c^{2},  \forall k, \label{eq:p1-2d} \\
	& \eqref{eq:rank1a}, \eqref{eq:rank1bconvex}, \notag
\end{align}
\end{subequations}
which is easily shown to be convex, and hence, $\mathbf{W}_k$ can be iteratively obtained by solving problem \eqref{eq:p1-2} based on updating $\tilde{\mathbf{w}}_k$ and  $\tilde{\mathbf{W}}_k$ in an iterative manner. However, due to the stringent requirement introduced by \eqref{eq:p1-2b} and \eqref{eq:rank1b}, it is generally non-trivial to directly obtain a feasible solution as an initial point. Alternatively, we can adopt the penalty SCA~\cite{wang2021intelligent} and introduce auxiliary variables $\bar{p}, \rho_k, \kappa$ to transform problem \eqref{eq:p1-2} into
\begin{subequations}  \label{eq:p1-3}
	\begin{align} 
&  \max_{\left\lbrace  \mathbf{w}_k, \mathbf{W}_k, \rho_k\right\rbrace_{k=1}^K, \kappa} I_r - \bar{p}  \kappa - \bar{p} \sum_{k=1}^{K} \rho_k\\
		s.t.~~& \tilde{I}_e \leq \epsilon + \kappa, \label{eq:p1-3b} \\
		& 	\operatorname{tr}(\mathbf{W}_k) - {\tilde{\mathbf{w}}_k}^H \tilde{\mathbf{w}}_k - 2\operatorname{Re}\left(\tilde{\mathbf{w}}_k^H \mathbf{w}_k \right) \leq \rho_k, \\
		&\eqref{eq:rank1a}, \eqref{eq:p1-power}, \eqref{eq:p1-2c},  \eqref{eq:p1-2d}, 
	\end{align}
\end{subequations}
where $\bar{p}$ and $\rho_k, \kappa$ denote the weight coefficient and the penalty terms, respectively. 
We present the proposed iterative algorithm in Algorithm 1.
\begin{algorithm} 
	\caption{\it: Proposed Iterative Algorithm for Handling \eqref{eq:p1-3}  } 
	\label{alg1} 
	\begin{algorithmic}
		\STATE Randomly set $ \{ \mathbf{w}_k^{(0)}, \mathbf{W}_k^{(0)} \} $, $\bar{p}^{(0)} = 10^{-3}$, $\lambda > 1$, $i = 0$;
		\REPEAT 
		\STATE $i \leftarrow i + 1$ ;
		\STATE $\tilde{\mathbf{w}}_k^{(i)}, \tilde{\mathbf{W}}_k^{(i)} \leftarrow \mathbf{w}_k^{(i-1)}, \mathbf{W}_k^{(i-1)}$;
		\STATE Solve problem \eqref{eq:p1-2} to obtain the optimal $\mathbf{w}_k^{(i)}, \mathbf{W}_k^{(i)}$;
			\STATE $\bar{p}^{(i)} \leftarrow \lambda \bar{p}^{(i-1)} $
		\UNTIL both $I_r^{(i)} - I_r^{(i-1)}$ and the penalty terms $\rho_k, \kappa$ are significantly small.   
	\end{algorithmic}
\end{algorithm}

%

\section{MI Gap Maximization with AN}

In this section, we consider utilizing \gls{AN} to assist secure sensing. We assume the instantaneous AN, $\mathbf{N}$, is shared with the legitimate receiver by the control center but remains unknown to the Eve. Firstly, we give the received signal at the legitimate receiver as
\begin{align}
	&\mathbf{Y}_{r} = \mathbf{H}_{r}(\bm{\theta}_r) \mathbf{X} + \mathbf{H}_{r}(\bm{\theta}_r) \mathbf{N} + \mathbf{Z}_{r}.
\end{align}
Then, the MI of the legitimate receiver can be expressed as

\begin{align}
	&~ I_r(\mathbf{Y}_{r}; \bm{\theta}_r | \mathbf{X}, \mathbf{N}) =  I_r(\mathbf{Y}_{r}; \mathbf{H}_{r} | \mathbf{X}, \mathbf{N}) \notag\\
	 & = h(\mathbf{Y}_{r}|  \mathbf{X}, \mathbf{N}) - h(\mathbf{Y}_{r} | \mathbf{H}_r,  \mathbf{X}, \mathbf{N}) \notag \\
	& = \log \det( \mathbf{I} + \sigma_{r}^{-2} (\tilde{\mathbf{X}} +\tilde{\mathbf{N}}) \mathbf{R}_{h_r} (\tilde{\mathbf{X}} + \tilde{\mathbf{N}})^H) \notag\\
	&=\log \det( \mathbf{I} + \sigma_{r}^{-2} (\tilde{\mathbf{X}} + \tilde{\mathbf{N}})^H(\tilde{\mathbf{X}} +\tilde{\mathbf{N}}) \mathbf{R}_{h_r} )  \notag \\
	&\overset{(a)}{\approx} \log \det\left(  \mathbf{I} + \sigma_{r}^{-2} L\mathbf{K}(\mathbf{I}_{N_r}\otimes\left({\mathbf{R}}^*_{\mathbf{X}} + {\mathbf{R}}^*_{\mathbf{N}}\right))\mathbf{K}^T\mathbf{R}_{h_r} \right)  \notag \\
	& =  \log \det\left(  \mathbf{I} + \sigma_{r}^{-2} L \mathbf{\Lambda}_r \sum_{i=1}^{N_r} \mathbf{P}_i (\mathbf{R}_\mathbf{X}^\ast + \mathbf{R}_\mathbf{N}^\ast ) \mathbf{P}_i^H \right) ,
\end{align}
where $\tilde{\mathbf{N}} = \mathbf{N}^T \otimes \mathbf{I}_{N_r}$ and ${\mathbf{R}}_{\mathbf{N}}=\frac{1}{L}\mathbf{N}\mathbf{N}^H$. 

On the other hand, as the Eve has no prior knowledge of AN, the received signal and the MI at the Eve can be given as 
\begin{align}
	&\mathbf{Y}_{e} = \mathbf{H}_{e}(\bm{\theta}_e) \mathbf{X} + \mathbf{H}_{e}(\bm{\theta}_e) \mathbf{N} + \mathbf{Z}_{e}, 
\end{align}
and
\begin{align}
	&~ I_e(\mathbf{Y}_{e}; \bm{\theta}_e | \mathbf{X}) =  I_e(\mathbf{Y}_{r}; \mathbf{H}_{e} | \mathbf{X}) = h(\mathbf{Y}_{e}|  \mathbf{X}) -h(\mathbf{Y}_{e} | \mathbf{H}_e,  \mathbf{X}),
\end{align}
respectively. Then, we have
\begin{align}
	& h(\mathbf{Y}_{e} | \mathbf{H}_e,  \mathbf{X}) \notag \\
	= & - \int_{\mathcal{X,Y,H}} f( \mathbf{X},  \mathbf{Y},  \mathbf{H}) \log f(\mathbf{Y} | \mathbf{X},  \mathbf{H}) d\mathbf{X}d\mathbf{Y}d \mathbf{H} \notag \\
	= &  - \int_{\mathcal{X,H}} f( \mathbf{X}, \mathbf{H}) \int_{\mathcal{Y}} f(\mathbf{Y} | \mathbf{X},  \mathbf{H})  \log f(\mathbf{Y} | \mathbf{X},  \mathbf{H}) d\mathbf{Y} d\mathbf{X} d \mathbf{H} \notag \\
	= & \!\! \int_{\mathcal{X,H}}\!\!\!\! \!\! \!\!  f( \mathbf{X}, \mathbf{H}) \log \left(\left( 2\pi e\right)^n\!\!  \det \left(L (\mathbf{R}_\mathbf{N}^\ast\!\! \otimes\!\mathbf{I}_{N_e}\! \! ) \mathbf{h}_e  \mathbf{h}_e^H \!\!  +\!\!  \sigma_e^2 \mathbf{I} \right)  \right)\!\!  d\mathbf{X} d \mathbf{H} \notag \\
	\approx & \frac{1}{J} \sum_{j=1}^{J} \log \det \left(2\pi e L (\mathbf{R}_\mathbf{N}^\ast \otimes \mathbf{I}_{N_e}) \mathbf{h}_{e,j}  \mathbf{h}_{e,j}^H  + \sigma_e^2 \mathbf{I} \right) , \notag \\
	= & \frac{1}{J} \sum_{j=1}^{J} \log \det \left(2\pi e L \mathbf{K}(\mathbf{I}_{N_e} \otimes  \mathbf{R}_\mathbf{N}^\ast) \mathbf{K}^T \mathbf{h}_{e,j}  \mathbf{h}_{e,j}^H  + \sigma_e^2 \mathbf{I} \right) \label{eq:Ie2}
\end{align}
where $\mathbf{h}_{e,j}(\bm{\theta})$ is the $j$-th sample of the random variable $\mathbf{h}_{e}(\bm{\theta})$. 
However, it's still difficult to derive the exact expression of $ h(\mathbf{Y}_{e}|  \mathbf{X})$ due to the additive non-Gaussian noise of $\mathbf{H}_{e}(\bm{\theta}_e) \mathbf{N} $. To deal with this, we can give the upper bound of $ h(\mathbf{Y}_{e}|  \mathbf{X})$ as
\begin{align}
	h(\mathbf{Y}_{e}|  \mathbf{X}) = \log \det \left( 2 \pi e \left( \sigma_{e}^2 \mathbf{I} + \mathbf{R}_{HN} \right) \right) ,
\end{align}
where $ \mathbf{R}_{HN}$ denotes the covariance of $\mathrm{vec}(\mathbf{HN})$, given as

\begin{align}
	\mathbf{R}_{HN} \!\!& = \mathbb{E}_{\mathbf{H},\mathbf{N}} \left[\left(\mathbf{I}_L \otimes \mathbf{H}_e \right) \mathrm{vec}(\mathbf{N}) \left( \mathrm{vec}(\mathbf{N})\right)^H  \left(\mathbf{I}_L \otimes \mathbf{H}_e \right)^H \right] \notag \\
	& = \mathbb{E}_{\mathbf{H}} \!\! \left[ \left(\mathbf{I}_L \otimes \mathbf{H}_e \right) \mathbb{E}_{\mathbf{H}}\!\! \left[  \mathrm{vec}(\mathbf{N}) \left( \mathrm{vec}(\mathbf{N})\right)^H \right] \!\! \left(\mathbf{I}_L \otimes \mathbf{H}_e \right)^H  \right] \notag \\
	& = \mathbb{E}_{\mathbf{H}} \left[ \left(\mathbf{I}_L \otimes \mathbf{H}_e \right) \left(\mathbf{I}_L \otimes \mathbf{R}_\mathbf{N} \right) \left(\mathbf{I}_L \otimes \mathbf{H}_e \right)^H  \right] \notag \\
	& \approx \frac{1}{J} \sum_{j=1}^{J}	\left(\mathbf{I}_L \otimes \mathbf{H}_{e,j} \right) \left(\mathbf{I}_L \otimes \mathbf{R}_\mathbf{N} \right) \left(\mathbf{I}_L \otimes \mathbf{H}_{e,j} \right)^H 
\end{align}
Combining with \eqref{eq:Ie2}, we give the approximate upper bound of $I_e$ as 
\begin{align}
\bar{I}_e =  & \log \! \det \!\! \left( \!\! \sigma_{e}^2 \mathbf{I} \!\! +\!\!  \frac{1}{J} \sum_{j=1}^{J}	\left(\mathbf{I}_L \otimes \mathbf{H}_{e,j} \right) \left(\mathbf{I}_L \otimes \mathbf{R}_\mathbf{N} \right) \left(\mathbf{I}_L \otimes \mathbf{H}_{e,j} \right)^H \!\!  \right)  \notag \\
& -  \frac{1}{J} \sum_{j=1}^{J} \log \det \left(L \mathbf{K}(\mathbf{I}_{N_e} \otimes  \mathbf{R}_\mathbf{N}^\ast) \mathbf{K}^T \mathbf{h}_{e,j}  \mathbf{h}_{e,j}^H  + \sigma_e^2 \mathbf{I} \right). \notag
\end{align}
Then, the problem of sensing security with AN can be formulated as
\begin{subequations} \label{eq:p2-1}
	\begin{align} 
		&\max_{\left\lbrace \mathbf{W}_k\right\rbrace_{k=1}^K, \mathbf{R}_\mathbf{N}}  I_r(\mathbf{Y}_{r}; \bm{\theta}_r | \mathbf{X}, \mathbf{N})	 \\
		\mathrm{s.t.}~&\bar{I}_e  \leq \epsilon, \label{eq:p2b} \\
		& \frac{\text{tr}\left(\mathbf{h}_k  \mathbf{W}_k \mathbf{h}_k^H \right)}{\sum\nolimits_{k=1, k\neq i}^{K} \text{tr}\left(\mathbf{h}_k  \mathbf{W}_i \mathbf{h}_k^H \right)+ \mathbf{h}_k \mathbf{R}_\mathbf{N} \mathbf{h}_k^H+\sigma_c^{2}} \geq \gamma_k, \forall k, \label{eq:p2sinr}\\
		&\text{tr}\left( \sum_{k=1}^K \mathbf{W}_k + \mathbf{R}_\mathbf{N} \right)   \leq P_0, \label{eq:p2d}\\
		& \eqref{eq:p1e}. \notag
	\end{align} %
\end{subequations} 
As previously discussed, the formulated problem is not convex due to the nonconvexity of \eqref{eq:p2b}. Hence, we introduce an auxiliary matrix $\mathbf{Q}$ and reformulate \eqref{eq:p2b} based on the Taylor expansion at $\tilde{\mathbf{Q}}$ as
\begin{subequations}
\begin{align}
	& \log \det(\tilde{\mathbf{Q}} ) + \mathrm{tr}(\tilde{\mathbf{Q}}  (\mathbf{Q} - \tilde{\mathbf{Q}} )) \label{eq:ie1} \\
	& -  \frac{1}{J} \sum_{j=1}^{J} \log \det \left(L \mathbf{K}(\mathbf{I}_{N_e} \otimes  \mathbf{R}_\mathbf{N}^\ast) \mathbf{K}^T \mathbf{h}_{e,j}  \mathbf{h}_{e,j}^H  + \sigma_e^2 \mathbf{I} \right) \leq \epsilon, \nonumber \\
	&    \mathbf{Q}\! -\!  \left( \! \sigma_{e}^2 \mathbf{I} \!\! +\!\!  \frac{1}{J} \sum_{j=1}^{J}	\left(\mathbf{I}_L \otimes \mathbf{H}_{e,j} \right) \left(\mathbf{I}_L \otimes \mathbf{R}_\mathbf{N} \right) \left(\mathbf{I}_L \otimes \mathbf{H}_{e,j} \right)^H \!  \right) \!  \succeq \! 0. \label{eq:ie2}
\end{align}
\end{subequations}
Therefore, a convex approximation of problem \eqref{eq:p2-1} is reformulated as
\begin{subequations}
	\begin{align}
&\max_{\left\lbrace \mathbf{w}_k, \mathbf{W}_k\right\rbrace_{k=1}^K, \mathbf{R}_\mathbf{N}}   I_r 	 \\
\mathrm{s.t.}~&\eqref{eq:rank1a}, \eqref{eq:rank1bconvex}, \eqref{eq:p1-2c}, \eqref{eq:p2sinr}, \eqref{eq:p2d}, \eqref{eq:ie1}, \eqref{eq:ie2} ,
	\end{align}
\end{subequations}
which can be solved in a similar iterative manner as in Algorithm 1.

\section{Simulation Results}

In this section, we provide numerical analysis to evaluate the performance of the proposed algorithms, including secure sensing de without AN and MI gap maximization with AN.  We consider a dual-functional BS transceiver equipped with $N_t$ = 6 transmit antennas for MIMO radar sensing and multi-user communication, serving $K = 3$ CUs where one of the CU serves as the sensing Eve. The legitimate radar receiver is equipped with $N_r = 2$ receive antennas. 
Unless stated otherwise, the available power budget $P_\text{max} = 30$ dBm and the frame length $L = 30$.  
We adopt the widely-used multi-user broadcasting beamforming design \cite[Sec. 7.6.1]{chi2017convex} as the baseline. 

\begin{figure}
	\centering
	\includegraphics[width=0.55\linewidth]{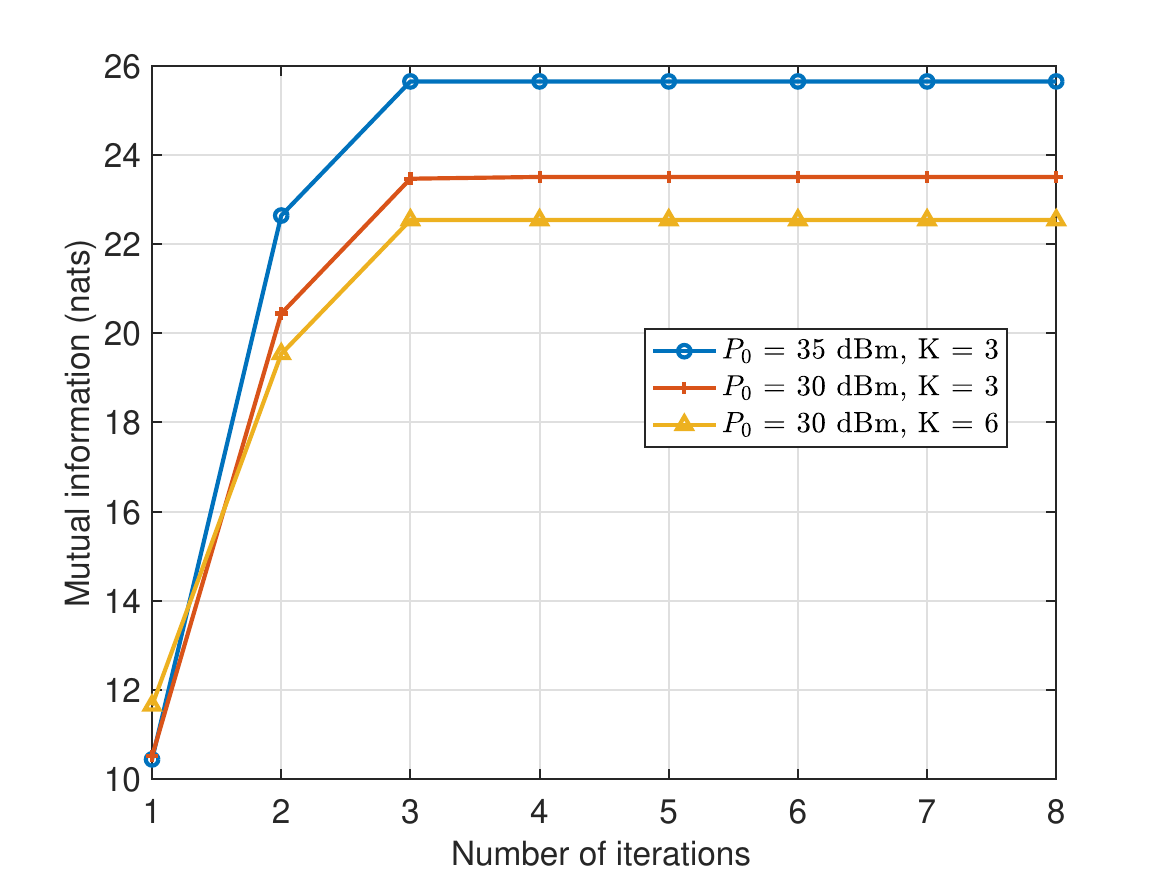}
	\caption{MI of the legitimate receiver versus the number of iterations. The limitation of Eve's MI is $\epsilon = 5$ nats and SINR threshold is $\gamma_k = 20$ dB.}
	\label{fig:fig1}
\end{figure}

Fig.~\ref{fig:fig1} firstly demonstrates the convergence behavior of the proposed method without the aid of \gls{AN}, under different power budgets and numbers of users. It can be seen that all three cases converge after 3 iterations, which verifies the fast convergence rate and the efficiency of our proposed alternating algorithm. With increasing $P_0$, the MI increases, since more power budget can be utilized to sense the target and satisfy the SINR constraints of multiple users. Moreover, increasing the number of CUs leads to a slight reduction in the sensing performance.

\begin{figure}
	\centering
	\includegraphics[width=0.55\linewidth]{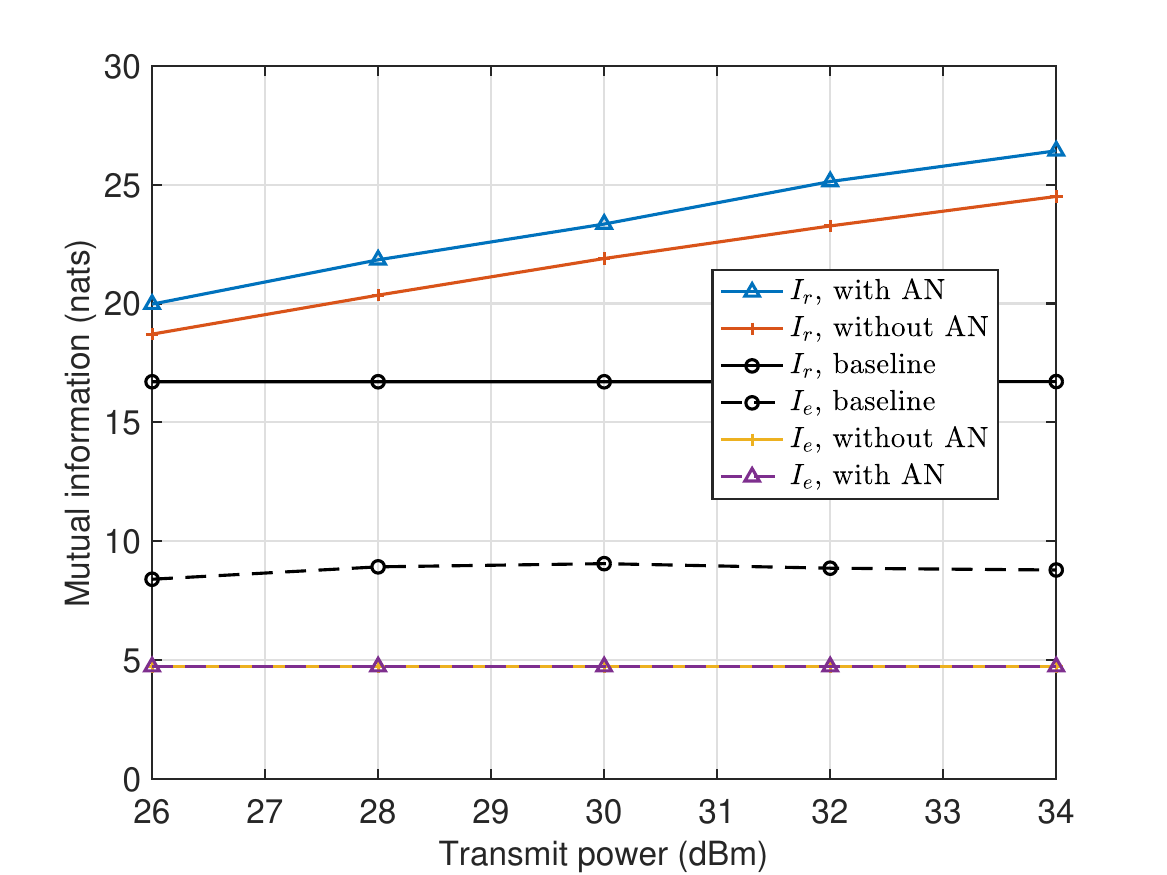}
	\caption{MI versus the transmit power budget, in both with and without AN-aided case, compared with the baseline. The limitation of the MI of the Eve is set as $\epsilon = 5$ nats and the SINR threshold is $\gamma_k = 28 dB$. }
	\label{fig:fig2}
\end{figure}

Fig.~\ref{fig:fig2} compares the MI performance with different transmitted power budgets, compared with the baseline scheme.
 It can be seen that the gap between $I_r$ and $I_e$ of the proposed algorithms is always larger than that of the baseline scheme, and over 2 times higher with the increase of transmit power budget. Moreover, it can be observed that the use of AN further improves the sensing MI of the legitimate receiver and enlarges the MI gap.
\vspace{-2mm}
\vspace{-2mm}
\section{Conclusion}
This paper addressed the sensing security issue for ISAC systems by beamforming design. The MI of radar sensing between the legitimate receiver was maximized while taking into account the MI of the potential \gls{Eve}, power budget, and SINR constraints. Additionally, we adopted AN to further guarantee secure sensing.
Simulation results demonstrated the effectiveness of the proposed methods in achieving a superior MI compared to the baseline scheme, with the AN further contributing to sensing security.

{\appendices
	
	\section*{Appendix A}
	Here, we provide the proof for the MI approximation for the \gls{Eve} in \eqref{eq:eveMI} based on a Taylor series expansion. First, defining $f(\mathbf{R}_\mathbf{X}) =  \log \det (\mathbf{I} +  \sigma_{e}^{-2} T \mathbf{\Lambda}_e \sum_{i=1}^{N_r} \mathbf{Q}_i \mathbf{R}_\mathbf{X}^\ast \mathbf{Q}_i^H)$ ,we have the Jacobian matrix of $f(\mathbf{R}_\mathbf{X})$ expressed as
	\begin{align}
		\mathrm{D}_{\mathbf{R}_\mathbf{X}^\ast} f(\mathbf{R_X}) =  \sigma_{e}^{-2} T \sum_{i=1}^{N_e}  \mathbf{Q}_i^H  \mathbf{M}^{-1} \mathbf{\Lambda}_e \mathbf{Q}_i.
	\end{align}
	Following \cite[Sec. 1.1.11]{chi2017convex} and \cite[Sec. 3.1]{zhang2017matrix}, the gradient of $f(\mathbf{R_X})$ can be given as
	\begin{align}
		\nabla_{\mathbf{R_X}} f(\mathbf{R_X}) & = 2 \nabla_{\mathbf{R^\ast_X}} f(\mathbf{R_X}) = 2 \sigma_{e}^{-2} T \sum_{i=1}^{N_e}  \mathbf{Q}_i^T   \mathbf{\Lambda}_e  (\mathbf{M}^{-1})^T \mathbf{Q}_i^\ast,
	\end{align}
	Then, an affine Taylor series approximation of $f(\mathbf{R_X})$ at $\mathbf{R_X} = \tilde{\mathbf{R}}_\mathbf{X} $ can be written as 
	\begin{align}
		& f(\mathbf{R_X})  \simeq f(\tilde{\mathbf{R}}_\mathbf{X} ) + \mathrm{tr}\left( \mathrm{Re}\left( \nabla f(\tilde{\mathbf{R}}_\mathbf{X} )^H \left( \mathbf{R_X} - \tilde{\mathbf{R}}_\mathbf{X} \right)  \right)  \right) \notag \\
		= & f(\tilde{\mathbf{R}}_\mathbf{X} ) + \mathrm{tr}\left( \mathrm{Re}\left( 2 \sigma_{e}^{-2} T  \sum_{i=1}^{N_e}  \mathbf{Q}_i^T  (\mathbf{M}^{-1})^\ast  \mathbf{\Lambda}_e  \mathbf{Q}_i^\ast \mathbf{R_X}  \right)  \right) \notag \\
		&-  \mathrm{tr}\left( \mathrm{Re}\left( 2 \sigma_{e}^{-2} T  \sum_{i=1}^{N_e}  \mathbf{Q}_i^T  (\mathbf{M}^{-1})^\ast  \mathbf{\Lambda}_e  \mathbf{Q}_i^\ast  \tilde{\mathbf{R}}_\mathbf{X}   \right)  \right)
	\end{align}

}

\bibliographystyle{IEEEtran}
\bibliography{sensingsecurity}

\vspace{12pt}

\newpage

\vfill

\end{document}